\documentclass{ws-ijmpe}

\begin{document}
\newcommand {\ee}{\end{equation}}
\newcommand {\bea}{\begin{eqnarray}}
\newcommand {\eea}{\end{eqnarray}}
\newcommand {\nn}{\nonumber \\}
\newcommand {\Tr}{{\rm Tr\,}}
\newcommand {\tr}{{\rm tr\,}}
\newcommand {\e}{{\rm e}}
\newcommand {\etal}{{\it et al.}}
\newcommand {\m}{\mu}
\newcommand {\n}{\nu}
\newcommand {\pl}{\partial}
\newcommand {\p} {\phi}
\newcommand {\vp}{\varphi}
\newcommand {\vpc}{\varphi_c}
\newcommand {\al}{\alpha}
\newcommand {\be}{\beta}
\newcommand {\ga}{\gamma}
\newcommand {\Ga}{\Gamma}
\newcommand {\ka}{\kappa}
\newcommand {\la}{\lambda}
\newcommand {\La}{\Lambda}
\newcommand {\si}{\sigma}
\newcommand {\Si}{\Sigma}
\newcommand {\Th}{\Theta}
\newcommand {\om}{\omega}
\newcommand {\Om}{\Omega}
\newcommand {\ep}{\epsilon}
\newcommand {\vep}{\varepsilon}
\newcommand {\na}{\nabla}
\newcommand {\del}  {\delta}
\newcommand {\Del}  {\Delta}
\newcommand {\mn}{{\mu\nu}}
\newcommand {\ls}   {{\lambda\sigma}}
\newcommand {\ab}   {{\alpha\beta}}
\newcommand {\gd}   {{\gamma\delta}}
\newcommand {\half}{ {\frac{1}{2}} }
\newcommand {\third}{ {\frac{1}{3}} }
\newcommand {\fourth} {\frac{1}{4} }
\newcommand {\sixth} {\frac{1}{6} }
\newcommand {\sqg} {\sqrt{g}}
\newcommand {\fg}  {\sqrt[4]{g}}
\newcommand {\invfg}  {\frac{1}{\sqrt[4]{g}}}
\newcommand {\sqZ} {\sqrt{Z}}
\newcommand {\gbar}{\bar{g}}
\newcommand {\sqk} {\sqrt{\kappa}}
\newcommand {\sqt} {\sqrt{t}}
\newcommand {\reg} {\frac{1}{\epsilon}}
\newcommand {\fpisq} {(4\pi)^2}
\newcommand {\Lcal}{{\cal L}}
\newcommand {\Ocal}{{\cal O}}
\newcommand {\Dcal}{{\cal D}}
\newcommand {\Ncal}{{\cal N}}
\newcommand {\Mcal}{{\cal M}}
\newcommand {\scal}{{\cal s}}
\newcommand {\Dvec}{{\hat D}}   
\newcommand {\dvec}{{\vec d}}
\newcommand {\Evec}{{\vec E}}
\newcommand {\Hvec}{{\vec H}}
\newcommand {\Vvec}{{\vec V}}
\newcommand {\Btil}{{\tilde B}}
\newcommand {\ctil}{{\tilde c}}
\newcommand {\Ftil}{{\tilde F}}
\newcommand {\Ktil}  {{\tilde K}}
\newcommand {\Ltil}  {{\tilde L}}
\newcommand {\mtil}{{\tilde m}}
\newcommand {\ttil} {{\tilde t}}
\newcommand {\Qtil}  {{\tilde Q}}
\newcommand {\Rtil}  {{\tilde R}}
\newcommand {\Stil}{{\tilde S}}
\newcommand {\Ztil}{{\tilde Z}}
\newcommand {\altil}{{\tilde \alpha}}
\newcommand {\betil}{{\tilde \beta}}
\newcommand {\etatil} {{\tilde \eta}}
\newcommand {\latil}{{\tilde \lambda}}
\newcommand {\ptil}{{\tilde \phi}}
\newcommand {\Ptil}{{\tilde \Phi}}
\newcommand {\natil} {{\tilde \nabla}}
\newcommand {\xitil} {{\tilde \xi}}
\newcommand {\Rhat}{{\hat R}}
\newcommand {\Shat}{{\hat S}}
\newcommand {\ehat}{{\hat e}}
\newcommand {\mhat}{{\hat m}}
\newcommand {\shat}{{\hat s}}
\newcommand {\Dhat}{{\hat D}}   
\newcommand {\Vhat}{{\hat V}}   
\newcommand {\xhat}{{\hat x}}
\newcommand {\Zhat}{{\hat Z}}
\newcommand {\Gahat}{{\hat \Gamma}}
\newcommand {\nah} {{\hat \nabla}}
\newcommand {\etahat} {{\hat \eta}}
\newcommand {\omhat} {{\hat \omega}}
\newcommand {\psihat} {{\hat \psi}}
\newcommand {\thhat} {{\hat \theta}}
\newcommand {\gh}  {{\hat g}}
\newcommand {\Kbar}  {{\bar K}}
\newcommand {\Lbar}  {{\bar L}}
\newcommand {\Qbar}  {{\bar Q}}
\newcommand {\labar}{{\bar \lambda}}
\newcommand {\cbar}{{\bar c}}
\newcommand {\bbar}{{\bar b}}
\newcommand {\Bbar}{{\bar B}}
\newcommand {\psibar}{{\bar \psi}}
\newcommand {\chibar}{{\bar \chi}}
\newcommand {\fbar}{{\bar 5}}
\newcommand {\bbartil}{{\tilde {\bar b}}}
\newcommand  {\vz}{{v_0}}
\newcommand  {\ez}{{e_0}}
\newcommand  {\mz}{{m_0}}
\newcommand {\intfx} {{\int d^4x}}
\newcommand {\inttx} {{\int d^2x}}
\newcommand {\change} {\leftrightarrow}
\newcommand {\ra} {\rightarrow}
\newcommand {\larrow} {\leftarrow}
\newcommand {\ul}   {\underline}
\newcommand {\pr}   {{\quad .}}
\newcommand {\com}  {{\quad ,}}
\newcommand {\q}    {\quad}
\newcommand {\qq}   {\quad\quad}
\newcommand {\qqq}   {\quad\quad\quad}
\newcommand {\qqqq}   {\quad\quad\quad\quad}
\newcommand {\qqqqq}   {\quad\quad\quad\quad\quad}
\newcommand {\qqqqqq}   {\quad\quad\quad\quad\quad\quad}
\newcommand {\qqqqqqq}   {\quad\quad\quad\quad\quad\quad\quad}
\newcommand {\lb}    {\linebreak}
\newcommand {\nl}    {\newline}

\newcommand {\vs}[1]  { \vspace*{#1 cm} }

\newcommand {\MPL}  {Mod.Phys.Lett.}
\newcommand {\NP}   {Nucl.Phys.}
\newcommand {\PL}   {Phys.Lett.}
\newcommand {\PR}   {Phys.Rev.}
\newcommand {\PRL}   {Phys.Rev.Lett.}
\newcommand {\CMP}  {Commun.Math.Phys.}
\newcommand {\JMP}  {Jour.Math.Phys.}
\newcommand {\AP}   {Ann.of Phys.}
\newcommand {\PTP}  {Prog.Theor.Phys.}
\newcommand {\NC}   {Nuovo Cim.}
\newcommand {\CQG}  {Class.Quantum.Grav.}


\font\smallr=cmr5
\def\ocirc#1{#1^{^{{\hbox{\smallr\llap{o}}}}}}
\def\ogamma{\ocirc{\gamma}{}}
\def\oM{{\buildrel {\hbox{\smallr{o}}} \over M}}
\def\osigma{\ocirc{\sigma}{}}

\def\overleftrightarrow#1{\vbox{\ialign{##\crcr
 $\leftrightarrow$\crcr\noalign{\kern-1pt\nointerlineskip}
 $\hfil\displaystyle{#1}\hfil$\crcr}}}
\def\overnab{{\overleftrightarrow\nabslash}}

\def\va{{a}}
\def\vb{{b}}
\def\vc{{c}}
\def\tilpsi{{\tilde\psi}}
\def\tbpsi{{\tilde{\bar\psi}}}

\def\delL{{\delta_{LL}}}
\def\delG{{\delta_{G}}}
\def\delc{{\delta_{cov}}}

\newcommand {\sqxx}  {\sqrt {x^2+1}}   
\newcommand {\gago}  {\gamma^5}
\newcommand {\Pp}  {P_+}
\newcommand {\Pm}  {P_-}
\newcommand {\GfMp}  {G^{5M}_+}
\newcommand {\GfMpm}  {G^{5M'}_-}
\newcommand {\GfMm}  {G^{5M}_-}
\newcommand {\Omp}  {\Omega_+}    
\newcommand {\Omm}  {\Omega_-}
\def\Aslash{{}\hbox{\hskip2pt\vtop
 {\baselineskip23pt\hbox{}\vskip-24pt\hbox{/}}
 \hskip-11.5pt $A$}}
\def\Rslash{{}\hbox{\hskip2pt\vtop
 {\baselineskip23pt\hbox{}\vskip-24pt\hbox{/}}
 \hskip-11.5pt $R$}}
\def\kslash{
{}\hbox       {\hskip2pt\vtop
                   {\baselineskip23pt\hbox{}\vskip-24pt\hbox{/}}
               \hskip-8.5pt $k$}
           }    
\def\qslash{
{}\hbox       {\hskip2pt\vtop
                   {\baselineskip23pt\hbox{}\vskip-24pt\hbox{/}}
               \hskip-8.5pt $q$}
           }    
\def\dslash{
{}\hbox       {\hskip2pt\vtop
                   {\baselineskip23pt\hbox{}\vskip-24pt\hbox{/}}
               \hskip-8.5pt $\partial$}
           }    
\def\dbslash{{}\hbox{\hskip2pt\vtop
 {\baselineskip23pt\hbox{}\vskip-24pt\hbox{$\backslash$}}
 \hskip-11.5pt $\partial$}}
\def\Kbslash{{}\hbox{\hskip2pt\vtop
 {\baselineskip23pt\hbox{}\vskip-24pt\hbox{$\backslash$}}
 \hskip-11.5pt $K$}}
\def\Ktilbslash{{}\hbox{\hskip2pt\vtop
 {\baselineskip23pt\hbox{}\vskip-24pt\hbox{$\backslash$}}
 \hskip-11.5pt ${\tilde K}$}}
\def\Ltilbslash{{}\hbox{\hskip2pt\vtop
 {\baselineskip23pt\hbox{}\vskip-24pt\hbox{$\backslash$}}
 \hskip-11.5pt ${\tilde L}$}}
\def\Qtilbslash{{}\hbox{\hskip2pt\vtop
 {\baselineskip23pt\hbox{}\vskip-24pt\hbox{$\backslash$}}
 \hskip-11.5pt ${\tilde Q}$}}
\def\Rtilbslash{{}\hbox{\hskip2pt\vtop
 {\baselineskip23pt\hbox{}\vskip-24pt\hbox{$\backslash$}}
 \hskip-11.5pt ${\tilde R}$}}
\def\Kbarbslash{{}\hbox{\hskip2pt\vtop
 {\baselineskip23pt\hbox{}\vskip-24pt\hbox{$\backslash$}}
 \hskip-11.5pt ${\bar K}$}}
\def\Lbarbslash{{}\hbox{\hskip2pt\vtop
 {\baselineskip23pt\hbox{}\vskip-24pt\hbox{$\backslash$}}
 \hskip-11.5pt ${\bar L}$}}
\def\Rbarbslash{{}\hbox{\hskip2pt\vtop
 {\baselineskip23pt\hbox{}\vskip-24pt\hbox{$\backslash$}}
 \hskip-11.5pt ${\bar R}$}}
\def\Qbarbslash{{}\hbox{\hskip2pt\vtop
 {\baselineskip23pt\hbox{}\vskip-24pt\hbox{$\backslash$}}
 \hskip-11.5pt ${\bar Q}$}}
\def\Acalbslash{{}\hbox{\hskip2pt\vtop
 {\baselineskip23pt\hbox{}\vskip-24pt\hbox{$\backslash$}}
 \hskip-11.5pt ${\cal A}$}}


\markboth{Shoichi Ichinose}{CP VIOLATION FROM A HIGHER DIMENSIONAL MODEL}

\catchline{}{}{}{}{}

\title{CP VIOLATION FROM A HIGHER DIMENSIONAL MODEL
}

\author{\footnotesize SHOICHI ICHINOSE
       }

\address{School of Food and Nutritional Sciences, University of Shizuoka\\
Yada 52-1, Shizuoka 422-8526, Japan
\\
first\_ichinose@u-shizuoka-ken.ac.jp}

\maketitle

\begin{history}
\received{(received date)}
\revised{(revised date)}
\end{history}

\begin{abstract}
It is shown that Randall-Sundrum model has the EDM term
which violates the CP-symmetry. The comparison with
the case of Kaluza-Klein theory is done. 
The chiral property, localization, anomaly phenomena
are examined. 
We evaluate the bulk quantum effect using the method of
the induced effective action. This is a new
origin of the CP-violation. 
\end{abstract}

\section{Introduction}
As the quantities to measure the extendedness of a particle, 
there are two important physical quantities:\ the magnetic dipole 
moment ({\bf $\m$},MDM) and the electric dipole moment (${\bf d}$,EDM).
\begin{eqnarray}
{\cal O}_1\ =\ \frac{e}{2m}F_{ab}{\bar \psi}\si^{ab}\psi\sim\ -{\bf \m}\cdot{\bf B}\com\nn
{\cal O}_2\ =\ \frac{e}{2m}F_{ab}{\bar \psi}\gago\si^{ab}\psi\sim\ -{\bf d}\cdot{\bf E}
\com
\label{Int1}
\end{eqnarray}
where ${\bf B}$ is the magnetic flux density vector and ${\bf E}$ is the electric field vector. 
Both quantities, ${\cal O}_1$ and ${\cal O}_2$, have the physical dimension of $M^5$
(higher-dimensional operators). Hence they do not appear in the starting Lagrangian, and 
usually appear only through the quantum effect. 
In particular EDM term, ${\cal O}_2$, violates the CP-symmetry. Because of this, the experimental efforts
to measure EDM, as well as MDM, has been made vigorously. The latest result of 
the upper bounds are
\begin{eqnarray}
d_n\q <\q 2.9\times 10^{-26}\ \mbox{e}\cdot\mbox{cm}\ (\mbox{\cite{nEDM0602}})\com\nn
d_e\q <\q 1.6\times 10^{-27}\ \mbox{e}\cdot\mbox{cm}\ (\mbox{\cite{eEDM02}})
\com
\label{Int2}
\end{eqnarray}
where $d_n$ is the neutron EDM and $d_e$ is the electron EDM. 
EDM term can appear in the standard model, but the estimated magnitude is 
$d_n=10^{-32}$e$\cdot$cm.\cite{Gavela82,KZ82,MCHP87}
This is far less than the present experimental bound. 
Hence the detection of EDM implies the new physics beyond the standard model. 
Thirring\cite{Thirr72}, very long time ago, showed EDM, as well as MDM, 
appear in a 4D model reduced from the 5 dimensional model, Kaluza-Klein theory. 
The expected magnitude is far less than the experimental upper bound 
even at the present time. Although the result is numerically not interesting, 
it has some qualitatively-interesting features such as the dual relation. 
In the recent development of the unified theories using the extra dimension(s), 
Randall-Sundrum\cite{RS9905,RS9906} model has become one of the strong candidates
for the realistic model. 
In the original papers by Randall and Sundrum, 
the comparison between the two models is made mainly in the
mass hierarchy. 
Here we focus on another aspect, that is, 
the {\it magnetic and electric dipole moment} terms. 

The Kaluza-Klein theory has the long history.\cite{Kal21,Klein26} 
It is characterized by {\it compactifying}
the extra manifold. In this procedure 
the radius of the compact manifold, $1/\mu$, is introduced
as the size parameter. 
On the other hand, in the Randall-Sundrum model\cite{RS9905,RS9906},
the {\it localized} configuration in the extra space is utilized, 
instead of compactifying the extra space. In this procedure
the size parameter, $1/k$, of the localization 
("thickness" of the wall) is introduced. 
Both approaches accomplish the dimensional reduction
by adjusting the size parameters.

The content of this paper is based on the result of Ref.\cite{KEK01,SI02PR}.  

\section{Kaluza-Klein Theory}
The 5D space-time manifold is described by the 4D coordinates
$x^a$ ($a=0,1,2,3$) and an {\it extra} coordinate $y$. We also use
the notation ($X^m$)=($x^a, y$), ($m=0,1,2,3,5$). 
With the general 5D metric $\gh_{mn}$,
\begin{eqnarray}
ds^2=\gh_{mn}(X)dX^mdX^n
\com
\label{KK1}
\end{eqnarray}
we assume the $S^1$ compactification condition for the
extra space.
\begin{eqnarray}
\gh_{mn}(x,y)=\gh_{mn}(x,y+\frac{2\pi}{\mu})
\com
\label{KK1b}
\end{eqnarray}
where $\mu^{-1}$ is the {\it radius} of the extra space circle.
We specify the form of the metric as
\begin{eqnarray}
ds^2=
g_{ab}(x)dx^adx^b+\e^{2\si(x)}(dy-fA_a(x)dx^a)^2
\com
\label{KK1c}
\end{eqnarray}
where $g_{ab}(x), A_a(x)$ and $\si(x)$ are all 4D quantities, namely,  
the 4D metric, the U(1) gauge field and the dilaton (Weyl scale) field, 
respectively. 
$f$ is a coupling constant. 
This specification is based on the
following additional assumptions.

\begin{description}
\item[1.]
$y$ is a {\it space} (not time) coordinate.
\item[2.]
The geometry is invariant under
the U(1) symmetry:\ 
$y\ra y+\La(x),\ A_a(x)\ra A_a(x)+\frac{1}{f}\pl_a\La$ .
\item[3.]
We ignore the massive modes in the KK-expansion of $\gh_{mn}(x,y)$.
\end{description}

We take the Cartan formalism to introduce Dirac fermions and 
to compute the geometric
quantities such as the connection and 
the Riemann curvature\cite{Thirr72}. 
The f\"{u}nf-bein $\ehat^\m_{~m}$ 
is given by
\begin{eqnarray}
(\ehat^\m_{~m})=\left(
\begin{array}{cc}
e^\al_{~a} & 0 \\
-fe^\si A_a   & e^\si
\end{array}
                 \right)
\label{KK7}
\end{eqnarray}
The connection is given as
\begin{eqnarray}
\omhat^\fbar_{~\fbar}
=-\half\pl_a\si~dx^a
\com\q 
\omhat^\fbar_{~\al}=-\omhat_\al^{~\fbar}
=\half\pl_a\si~e_\al^{~a}\thhat^\fbar -\frac{f}{2}e^\si F_\ab\thhat^\be\com\nn
\omhat^\al_{~\be}=\om^\al_{~\be}
+\frac{f}{2}e^\si F^\al_{~\be}~\thhat^\fbar\com\ 
F_{ab}=\pl_aA_b-\pl_bA_a\com\q
F_\ab\equiv e_\al^{~a}e_\be^{~b}F_{ab}
\com
\label{KK9}
\end{eqnarray}
where $\om^\al_{~\be}$ is the 4D connection. 
The 5D Riemann scalar curvature
$\Rhat^{\m~\n}_{~\n~\m}$ can be decomposed as
\begin{eqnarray}
\Rhat=R+\frac{f^2}{4}e^{2\si}F^\ab F_\ab
+\half\pl_a\si\pl_b\si g^{ab}+\half\na^2\si
\com
\label{KK12}
\end{eqnarray}
which shows the theory of {\it gravity}, 
{\it electro-magnetism} and the dilaton in 
the 4D world. 

We consider the simple case $\si=0$ for the present purpose. 
\section{Fermions in Kaluza-Klein Theory}
The 5D Dirac equation is generally given by
\begin{eqnarray}
\left\{
\ga^\m\ehat_\m^{~m}\frac{\pl}{\pl X^m}
+\frac{1}{8}(\omhat^\si)_\mn\ga_\si [\ga^\m,\ga^\n]+\mhat
\right\}\psihat=0
\pr
\label{ferKK1}
\end{eqnarray}
$\mhat$ is the mass parameter of the 5D fermion 
($-\infty<\mhat<\infty$).
$(\omhat^\si)_\mn$ is the spin connection.  
For simplicity we switch off the 4D gravity: 
$e^\al_{~a}\ra\del^\al_a\ ,\ \om^\al_{~\be}\ra 0$. 
\begin{eqnarray}
\left\{
\ga^a(\pl_a+fA_a\pl_5)+\ga^5\pl_5
-\frac{f}{16}F_{ab}\ga^5[\ga^a,\ga^b]+\mhat
\right\}\psihat=0
\com
\label{ferKK2}
\end{eqnarray}
where $\pl_5=\pl/\pl y$. \nl
\nl
(A) Charged Fermion\nl
Corresponding to a {\it massive} mode in the {\it 4D reduction}, 
we consider the following form for a charged fermion.
\begin{eqnarray}
\psihat(x,y)=\e^{i(\phi\gago+\m y)}\psi(x)
\pr
\label{ferKK3}
\end{eqnarray}
Here we regard the charged fermion as a KK-{\it massive} mode.
The {\it phase parameter} $\phi$ is chosen as
\begin{eqnarray}
(i\gago\m+\mhat)\e^{2i\phi\gago}
=\sqrt{{\mhat}^2+\m^2}\equiv M\pr\nn
(i)\ \mhat\neq 0\ :\q\tan 2\phi=-\frac{\m}{\mhat}\ ,\nn
-\frac{\pi}{2}<2\phi\leq 0 \ \ \mbox{for}\ \mhat>0;\ \ 
\pi\leq 2\phi< \frac{3\pi}{2}\ \ \mbox{for}\ \mhat<0
\pr\nn
(ii)\ \mhat=0\ :\q 2\phi=-\frac{\pi}{2}\pr
\label{ferKK4}
\end{eqnarray}
See the lower half region of Fig.1.
\begin{figure}[th]
\centerline{
\psfig{file=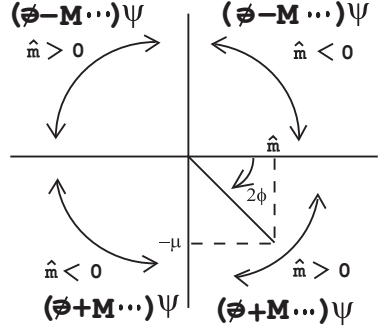,width=5cm}
           }
\vspace*{8pt}
   \caption{
Fig.1\ The angle parameter $\phi$ which defines the 4D charged
fermion in the KK dimensional reduction (\ref{ferKK3}).
$\mhat$ is the 5D fermion mass parameter, $\m^{-1}$ is the
size of the extra compact manifold. 
$-\pi/2 <2\phi\leq 0\ \mbox{for}\ \mhat>0;\  
\pi \leq 2\phi< 3\pi/2\ \mbox{for}\ \mhat<0;\ 
2\phi=-\pi/2\ \mbox{for}\ \mhat=0$.
The upper half region gives the same 4D fermion action
as the lower half region by the transformation $\psi\ra\gago\psi$.
           }
\end{figure}

Then (\ref{ferKK2}) reduces to
\begin{eqnarray}
\left\{
\ga^a(\pl_a+ieA_a)+M
-\frac{1}{16M}(\mhat\frac{e}{\m}-ie\gago)F_{ab}\gago [\ga^a,\ga^b]
\right\}\psi=0
\com
\label{ferKK5}
\end{eqnarray}
where $e\equiv f\m$ is the {\it electric coupling constant}.
$M$ is identified as a 4D fermion mass.
We notice, in this expression, 
the EDM and  the MDM naturally appear\cite{Thirr72}.
We consider the following two limits:\ 
\begin{description}
\item[(i) CP-preserved limit [Small radius limit, 4D limit]]\nl
$\mhat/\m \ra \pm 0$,$2\phi\ra 
-(\pi/2-0)$ [$3\pi/2-0$] for upper [lower] sign;\ 
\item[(ii) CP-extremely-violated limit [Large radius limit, 5D limit]]\nl
$\mhat/\m \ra \pm\infty$, $2\phi\ra -0$ [$\pi+0$] for upper [lower] sign.
\end{description}
From these results, we see CP-violation comes from the presence
of the connection term in (\ref{ferKK1}) (5 dimensionality itself), 
while the introduction of the phase parameter $\phi$ in (\ref{ferKK3})
makes MDM appear. (cf. Kobayashi-Maskawa's CP-phase. )
\nl
\nl
(B) Neutral Fermion\nl
We regard the neutral fermion as the {\it zero} mode of the KK-expansion.
\begin{eqnarray}
\psihat(x,y)=\psi(x)\com\nn
\left\{
\ga^a\pl_a
-\frac{f}{16}F_{ab}\ga^5[\ga^a,\ga^b]+\mhat
\right\}\psi(x)=0\pr
\label{ferKK6b}
\end{eqnarray}
Only the EDM term appears. 
Although the fermion has no charge, the
dipole moment appears. This case is similar to the limit (ii).\nl
\nl

Let us do the order estimation. From the reduction
\begin{eqnarray}
S=-\frac{1}{2G_5}\int d^5X\sqrt{-\gh}\Rhat=
-\frac{1}{2G_5}\frac{2\pi}{\m}\int d^4x 
\sqrt{-g}(R+\frac{f^2}{4}F^\ab F_\ab+\cdots)
\com
\label{ferKK7}
\end{eqnarray}
we know
\begin{eqnarray}
\frac{1}{G_5\m}\sim \frac{1}{G}\com\q
\frac{f^2}{G_5\m}\sim 1
\com
\label{ferKK8}
\end{eqnarray}
where $G$ is the (4D) gravitational constant. 
This gives $f\sim \sqrt{G}=10^{-19}\mbox{GeV}^{-1}$.
On the other hand, we know $e=\m f\sim 10^{-1}$.
Hence we obtain $\m\sim 10^{-1}f^{-1}\sim 10^{18}\mbox{GeV}$.
We originally have four parameters, $\mhat, \m, G_5$ and $f$. 
Among them we have the two relations (\ref{ferKK8}) from the
observation. We have another restriction among them from 
the present theoretical knowledge.
The most natural interpretation of the parameters is
that $\m^{-1}$ is the {\it infrared regularization}, $\mhat$ is
the energy scale of this 5D KK system. Then the validity
of the 5D {\it classical} treatment requires that
the 5D Planck mass $\gg |\mhat|$: 
\begin{eqnarray}
\frac{1}{\sqrt[3]{G_5}}\sim\sqrt[3]{100}\m
\gg |\mhat|\com
\label{ferKK8b}
\end{eqnarray}
and this reduces to, through the previous parameter
relations and values, $|\mhat|\ll 10^{19}$GeV. 

Now we consider the following three cases
in order to evaluate the EDM and MDM couplings.
(Note that the Planck length $\sim\ 10^{-32}$cm.)
\begin{description}
\item[(0) $|\mhat|\sim\m$]
We can estimate 
the electric and magnetic couplings as
\begin{eqnarray}
\frac{|\mhat|}{M\m}e
\sim 10^{-32}~\mbox{e cm}\com\q
\frac{e}{M}
\sim 10^{-32}~\frac{\mbox{cm}}{\mbox{e}}/\mu_0
\com
\label{ferKK9}
\end{eqnarray}
where $\mu_0$ is the permeability constant.
\footnote{
Bohr magneton e$\hbar/$2m$_e$=$1.7\times 10^{-12}$ ($\hbar\cdot$cm/e)/$\m_0$.
} 
Both electric and magnetic moment terms
equally appear.
In this case, however, the theoretical restriction (\ref{ferKK8b})
is not so well satisfied.
\item[(i) $|\mhat|\ll\m$(small radius, 4D limit), 
CP-preserved limit]
We can estimate as
\begin{eqnarray}
\frac{\mhat}{M\m}e\sim\frac{\mhat}{\m}\times 10^{-32}~\mbox{e cm}\ ,\ 
\frac{e}{M}\sim\frac{e}{\m}\sim 10^{-32}~\frac{\mbox{cm}}{\mbox{e}}/\mu_0\ 
\pr
\label{ferKK10}
\end{eqnarray}
In this case the EDM coupling is suppressed by the 
factor of the mass parameter ratio $\frac{|\mhat|}{\m}$($\ll 1$).
The theoretical restriction (\ref{ferKK8b})
is satisfied, hence this parameter region is well controlled
theoretically.
\item[(ii) $|\mhat|\gg\m$(large radius, 5D limit), 
CP-extremely-violated limit]
We can estimate as
\begin{eqnarray}
\frac{|\mhat|}{M\m}e\sim\frac{e}{\m}\sim 10^{-32}~\mbox{e cm}\com\q
\frac{e}{M}\sim\frac{\m}{|\mhat|}\times 10^{-32}~\frac{\mbox{cm}}{\mbox{e}}/\mu_0\ 
\pr
\label{ferKK11}
\end{eqnarray}
In this case the MDM coupling is suppressed by the  
factor $\frac{\m}{|\mhat|}$($\ll 1$).
The theoretical restriction (\ref{ferKK8b}), however,
is not satisfied. This implies the 5D {\it quantum} effect can not be
negligible in this parameter region.
\end{description}
We note that the ratio of 
the two massive parameters, 
(the radius of the extra space)$^{-1}$ $\m$ and 
the 5D fermion mass $\mhat$, controls
the {\it dual} aspect (electric versus magnetic) of the theory.
This point will be compared with the RS case later. 

As for the EDM,
all cases are {\it far below} the experimental upper bound
described in (\ref{Int2}). 
As for the MDM,  
we know, (from the formula:\ 
$e/2m$,) the order
of the observed values are
$10^{-11}$ ~$\frac{\mbox{cm}}{\mbox{e}}/\mu_0$ for the electron, 
$10^{-14}$ ~$\frac{\mbox{cm}}{\mbox{e}}/\mu_0$ for the proton,
$10^{-16}$ ~$\frac{\mbox{cm}}{\mbox{e}}/\mu_0$  for the top quark.
The prediction of 5D KK theory is {\it superweak} compared with these values. 
Hence the present model is viable 
but quantitatively not so attractive.
We are, at present, 
content with the qualitatively interesting points. 

Thirring\cite{Thirr72} showed that 
the CP-violating term (EDM term) appears
not because the discrete symmetries
(charge conjugation, parity and time reversal) 
do not exist in 
the Dirac equation (\ref{ferKK5}), 
but because they
appear in the form which differs from the ordinary one.

\section{Fermions in Randall-Sundrum Theory
}
We consider the following 5D space-time geometry\cite{RS9905,RS9906}.
\begin{eqnarray}
ds^2=\e^{-2\si(y)}\eta_{ab}dx^adx^b+dy^2
=\gh_{mn}dX^mdX^n\com\nn
-\infty<y<+\infty\ ,\ -\infty<x^a<+\infty
\com
\label{RS1}
\end{eqnarray}
where $\si(y)$ is a "scale factor" field.
$(\eta_{ab})=\mbox{diag}(-1,1,1,1)$.
When the geometry is AdS$_5$, $\si(y)=c|y|, c>0$. 

The f\"{u}nf-bein 
$\ehat^\m_{~m}$ is given by
\begin{eqnarray}
(\ehat^\m_{~m})=\left(
\begin{array}{cc}
e^{-\si}\eta^\al_{~a} & 0 \\
0   & 1
\end{array}
                 \right)
                 \pr
\label{RS4}
\end{eqnarray}
We obtain the connection 1-form
$\omhat^\m_{~\n}$ as
\begin{eqnarray}
\omhat^\fbar_{~\fbar}=0\com\q 
\omhat^\al_{~\fbar}=-\omhat_\fbar^{~\al}=-\si'\theta^\al\com\q
\omhat^\fbar_{~\al}=-\omhat_\al^{~\fbar}=\si'\theta_\al\com\q
\omhat^\al_{~\be}=0
\com
\label{RS5}
\end{eqnarray}
where $\si'=\frac{d\si}{dy}$.

The 5D Dirac Lagrangian in the RS theory is given, 
from (\ref{ferKK1}) and the results of Sec.4, as
\begin{eqnarray}
\sqrt{-\gh}(\Lcal^{Dirac}+i\mhat(y){\bar \psihat}\psihat)=
i\e^{-\frac{3}{2}\si}{\bar \psihat}\{
\ga^a\pl_a-2\e^{-\si}(\fourth\si'-\half\pl_y)\gago
                      +\mhat(y) e^{-\si}  \}(\e^{-\frac{3}{2}\si}\psihat),
\label{ferRS1}
\end{eqnarray}
where $\mhat$ is the 5D fermion mass $-\infty<\mhat<+\infty$.
For the later use, we here allow $\mhat$ to have the $y$-dependence :\ 
$\mhat=\mhat(y)$. 

Let us do the dimensional reduction from 5D to 4D
\cite{KS00,GN99,CHNOY99}.
We take the following form of expansion.
\begin{eqnarray}
\psihat(x,y)=\sum_n (\psi^n_L(x)\xi_n(y)+\psi^n_R(x)\eta_n(y)),\nn
\gago\psi_L(x)=-\psi_L(x)\com\q 
\gago\psi_R(x)=+\psi_R(x)
\com
\label{ferRS2}
\end{eqnarray}
where $\{\xi_n(y),\eta_n(y)\}$ is a complete set of some
eigenfunctions to be determined. 
For simplicity, we consider
the "5D-{\it parity}" even case for $\psihat(x,y)$.
\begin{eqnarray}
\gago\psihat(x,-y)=+\psihat(x,y)
\pr
\label{ferRS3}
\end{eqnarray}
This requires $\xi_n(y)$ to be an {\it odd} function and
$\eta_n(y)$ to be an {\it even} function with respect to
the {\it $Z_2$ transformation}:\ $y\change -y$. 
\begin{eqnarray}
\xi_n(-y)=-\xi_n(y)\com\q \eta_n(-y)=+\eta_n(y)
\pr
\label{ferRS4}
\end{eqnarray}
From these we get the following important boundary
conditions,
\begin{eqnarray}
\xi_n(0)=0\q\mbox{(Dirichlet)}\com\q 
\pl_y\eta_n|_{y=0}=0\q\mbox{(Neumann)}
\com
\label{ferRS4b}
\end{eqnarray}
when $\xi_n(y)$ and $\pl_y\eta_n(y)$ are continuous
at $y=0$. 
The Lagrangian reduces to
\begin{eqnarray}
\sqrt{-\gh}(\Lcal^{Dirac}+i\mhat(y){\bar \psihat}\psihat)
=i\sum_m(\psibar^m_L\xitil_m(y)+\psibar^m_R\etatil_m(y))\times\nn
\sum_n\{
\ga^a\pl_a\psi^n_L\xitil_n-
\e^{-\si}(\frac{\si'}{2}-\pl_y)(-\psi^n_L)\xitil_n+
\ga^a\pl_a\psi^n_R\etatil_n-
\e^{-\si}(\frac{\si'}{2}-\pl_y)\psi^n_R\etatil_n      \nn
+\e^{-\si}\mhat(y) (\psi^n_L\xitil_n+\psi^n_R\etatil_n)
  \}
\com
\label{ferRS5}
\end{eqnarray}
where we define $\xitil_n\equiv \e^{-\frac{3}{2}\si}\xi_n$ and
$\etatil_n\equiv \e^{-\frac{3}{2}\si}\eta_n$.

We now take the set of eigenfunctions $\{\xitil_n,\etatil_n\}$ as
\begin{eqnarray}
\e^{-\si}(\frac{\si'}{2}-\pl_y)\xitil_n
+\e^{-\si}\mhat(y)\xitil_n=m_n\etatil_n\ ,\nn
-\e^{-\si}(\frac{\si'}{2}-\pl_y)\etatil_n
+\e^{-\si}\mhat(y)\etatil_n=m_n\xitil_n
,
\label{ferRS6}
\end{eqnarray}
which are orthnormalized as
\begin{eqnarray}
\int^\infty_{-\infty}dy~\xitil_n(y)\xitil_m(y)=
\int^\infty_{-\infty}dy~\etatil_n(y)\etatil_m(y)=\del_{nm}\com\ 
\int^\infty_{-\infty}dy~\xitil_n(y)\etatil_m(y)=0
\ .
\label{ferRS7}
\end{eqnarray}
Then the 5D action (\ref{ferRS1}) finally reduces to the sum of 4D {\it free}
fermions.
\begin{eqnarray}
\int d^5X\sqrt{-\gh}(\Lcal^{Dirac}+i\mhat(y){\bar \psihat}\psihat)=\nn
i\int d^4x\sum_n\{
\psibar^n_L(\ga^a\pl_a\psi^n_L+m_n\psi^n_R)+
\psibar^n_R(\ga^a\pl_a\psi^n_R+m_n\psi^n_L)
                \}
.
\label{ferRS8}
\end{eqnarray}
The information of this fermion dynamics is now in 
the set of the eigen values
$\{m_n\}$ determined by (\ref{ferRS6}). 

From the coupled equation (\ref{ferRS6}) with respect to
$\xitil_n$ and $\etatil_n$, we get
the differential equation for $\xitil_n$ as
\begin{eqnarray}
\e^{-2\si}[\frac{\si''}{2}-\frac{3}{4}{\si'}^2
+2\si'\pl_y-\mhat(y)\si'+{\mhat(y)}^2
-{\pl_y}^2+{\mhat(y)}']\xitil_n={m_n}^2\xitil_n
\pr
\label{ferRS9}
\end{eqnarray}
For simplicity, we consider the {\it thin wall limit}:
\begin{eqnarray}
\si(y)=\om|y| \com\q \si'(y)=\om\ep(y)\com\q
\si''(y)=2\om\del(y)\com\nn
\q\q\q\q\q\mhat(y)=\mtil\ep(y)\com\q\mhat'(y)=2\mtil\del(y)
\com
\label{ferRS10}
\end{eqnarray}
where $\om(>0)$ and $\mtil(>0)$ are some constants.
$\ep(y)$ is the sign function:\ 
$\ep(y)=1$ for $y>0$ and $\ep(y)=-1$ for $y<0$
\ ($\ep'(y)=2\del(y)$). 
In this
limit, the equation (\ref{ferRS9}) can be explicitly
solved. 
\begin{eqnarray}
\e^{-2\om|y|}[(\om+2\mtil)\del(y)-\frac{3}{4}\om^2
+2\om\ep(y)\pl_y-\mtil\om+\mtil^2
-{\pl_y}^2]\xitil_n={m_n}^2\xitil_n
\com
\label{ferRS11}
\end{eqnarray}
where ${\ep(y)}^2=1$ is used. The presence of $\del(y)$ indicates 
a singularity of the solution at $y=0$. First 
let us see the solution in the region $y>0$. 
In terms of a new coordinate $z\equiv\frac{1}{\om}\e^{\om y}$,
the above equation reduces to the {\it Bessel differential
equation}.  
\begin{eqnarray}
\{{\pl_z}^2-\frac{1}{z}\pl_z+\frac{1-\n^2}{z^2}
+{m_n}^2\}\xitil_n=0\com\q
\n=|\frac{\mtil}{\om}-\half|
\pr
\label{ferRS12}
\end{eqnarray}
The solution $\xitil_n$ is obtained as
\begin{eqnarray}
\xitil_n(y)=\frac{1}{(\om z)^{3/2}}\xi_n(z)
=z\{ J_\n(m_nz)+c_nN_\n(m_nz)\}\com\q
c_n=-\frac{J_\n(m_n/\om)}{N_\n(m_n/\om)}
\com
\label{ferRS13}
\end{eqnarray}
where $c_n$ is determined by the Dirichlet boundary
condition (\ref{ferRS4b}). $J_\n(z)$ and $N_\n(z)$
are two independent Bessel functions. 
As for the solution valid for the full region $-\infty < y < \infty$,
we can obtain, taking into accout the singularity at $y=0$ and 
the odd property (\ref{ferRS4}), as 
\begin{eqnarray}
\xitil_n(y)=\mbox{e}^{-3\om|y|/2}\xi_n(y)
=\frac{\ep(y)}{\om}\mbox{e}^{\om|y|}\{ J_3(\frac{m_n}{\om}\mbox{e}^{\om|y|})
+c_nN_3(\frac{m_n}{\om}\mbox{e}^{\om|y|})\}\com\nn
c_n=-\frac{J_3(m_n/\om)}{N_3(m_n/\om)}
\com
\label{ferRS14}
\end{eqnarray}
where the parameter $\nu$ in (\ref{ferRS13})
is fixed to be 3. ($\mtil=-\frac{5}{2}\om$)
With the above
explicit solution of $\xitil_n$, we can obtain $\etatil_n$
using the first equation of (\ref{ferRS6}).
Another boundary condition (Neumann) on $\eta_n$ (\ref{ferRS4b})
gives us the set $\{m_n\}$ as the zeros of some combination
of Bessel functions.

The importance of the 5D mass "function" $\mhat(y)$ is now clear.
In the next section, we explain  its origin in the {\it bulk} field theory.
Nature requires the {\it Yukawa interaction} 
between the 5D fermion and the 5D Higgs\cite{BG99}.

\section{Bulk Higgs Mechanism and Massless Chiral Fermion Localization}
One of the most important characters of the brane world model
is the {\it massless chiral fermion localization}. It is phenomenologically
attractive because the smallness of the quark and lepton masses, 
compared with the Planck mass, 
could be naturally explained. Theoretically it is also necessary
as the dimensional reduction mechanism. The feature comes from
the $Z_2$ ($y\change -y$) properties of the system. 
The most natural way to introduce the properties
is to use the
{\it Higgs mechanism} in the bulk world. 
\footnote
{In the case of the flat space-time, 
Rubakov and Shaposhnikov\cite{RS83}
proposed a domain wall model
caused by the bulk Higgs potential.
}

Let us examine the case the fermion system has
the Yukawa coupling with the bulk Higgs field.
\begin{eqnarray}
\sqrt{-\gh}\Lcal=\sqrt{-\gh}(\Lcal^{Dirac}+\Lcal^Y)\com\q
\Lcal^Y=ig_Y{\bar \psihat}\psihat\Phi
\com
\label{yuka1}
\end{eqnarray}
where the Higgs field $\Phi$ is the 5D(bulk) scalar field
and $g_Y$ is the Yukawa coupling.
We assume that the Higgs field, besides 
the "scale factor" field $\si(y)$, 
is some background given by 
the (classical) field equation of
the 5D gravity-Higgs system.
\begin{eqnarray}
\sqrt{-\gh}(\Lcal^{grav}+\Lcal^S)\com\q
\Lcal^{grav}=\frac{-1}{2G_5}{\hat R}\com\q
\Lcal^{S}=-\half\na_m\Phi\na^m\Phi-V(\Phi)
\com
\label{yuka2}
\end{eqnarray}
where 
$V(\Phi)$ is the ordinary Higgs potential.
In ref.\cite{SI0003,SI0107}, it is shown that the above
gravity-Higgs system has a {\it stable} kink (domain wall)
solution for the case $\Phi=\Phi(y)$. 
In the IR asymptotic region far from the wall,
$\si'(y)$ and $\Phi(y)$ behave as
\begin{eqnarray}
\si'(y)=\left\{\begin{array}{c}+\om,\ \ ky\ra +\infty\\
                               -\om,\ \ ky\ra -\infty
                \end{array}\right.\com\q
\Phi(y)=\left\{
\begin{array}{c}
+\vz,\ \ ky\ra +\infty \\
-\vz,\ \ ky\ra -\infty

\end{array}
                 \right.                 
\com\label{yuka3}
\end{eqnarray}
where $k$(the inverse of thickness), $\om$(brane tension)
and $\vz$(5D Higgs vacuum expectation value) are some positive
constants expressed by a free parameter, the vacuum parameters
and the 5D gravitational constant. Near the origin of the extra axis
($k|y|\ll 1$), they behave as
\begin{eqnarray}
\si'(y)=\om\tanh (ky)
\com\q
\Phi(y)=\vz\tanh (ky)
\pr
\label{yuka4}
\end{eqnarray}

The dimensional reduction to 4D is performed by
taking the the {\it thin wall} limit $k\ra \infty$, 
which is precisely defined as
\begin{eqnarray}
k\gg \frac{1}{r_c}
\pr
\label{yuka4b}
\end{eqnarray}
where $r_c$ is the {\it infrared cutoff} 
of the extra axis ($-r_c<y<r_c$). (See ref.\cite{SI0003}.)
In this limit, above quantities behave as 
$\si'(y)\ra \om\ep(y),\ \Phi(y)\ra \vz\ep(y)$.  
All dimensional parameters are
a) ${G_5}^{-1/3}$:\ 5D Planck mass;\ 
b) $|\mhat|=g_Y\vz$:\ 5D fermion mass;\ 
c) $k^{-1}$:\ thickness of the domain;\ 
d) $r_c$:\ Infrared regularization of the extra axis.
Among them  there exists a theoretical restriction
from the requirement:\ 
5D {\it classical} treatment works well.
\begin{eqnarray}
\frac{1}{\sqrt[3]{G_5}}\gg k\com
\label{yuka4c}
\end{eqnarray}

The 5D Dirac equation of (\ref{yuka1})
 is given by (cf. eq.(\ref{ferRS1})),
\begin{eqnarray}
i\e^{\si}\{
\ga^a\pl_a-2\e^{-\si}(\si'-\half\pl_y)\gago+g_Y\e^{-\si}\Phi
                           \}\psihat=0
\pr
\label{yuka5}
\end{eqnarray}
This is just the lagragian of (\ref{ferRS1})
with $\mhat(y)=g_Y\Phi(y)$. 
Let us examine a solution of the left chirality zero mode.
\begin{eqnarray}
\psihat(x,y)=\psi^0_L(x)\eta(y)\com\q
\gago\psi^0_L=-\psi^0_L\com\q
\ga^a\pl_a\psi^0_L=0
\pr
\label{yuka6}
\end{eqnarray}
The equation (\ref{yuka5}) reduces to
\begin{eqnarray}
\pl_y\eta=(2\si'+g_Y\Phi(y))\eta
\pr
\label{yuka7}
\end{eqnarray}
In the IR asymptotic region($k|y|\gg 1$), 
the solution behaves as
\begin{eqnarray}
\eta(y)=\mbox{const}\times \e^{(g_Y\vz+2\om)|y|}
\com
\label{yuka8}
\end{eqnarray}
which shows the {\it exponentially damping} for the case:\ 
$g_Y\vz+2\om<0$\cite{BG99}. 
This is called
{\it massless chiral fermion localization}. Near the origin
of the extra axis($k|y|\ll 1$), $\eta(y)$ behaves as
\begin{eqnarray}
\eta(y)=\mbox{const}\times \e^{\frac{k}{2}(g_Y\vz+2\om)y^2}
\com
\label{yuka9}
\end{eqnarray}
which shows the {\it Gaussian damping}. The behavior is shown
in Fig.2. It shows the {\it regular} property of the solution.
For the right chirality zero-mode, we can show the same behavior
using the anti-kink solution instead of the kink solution (\ref{yuka3}). 
\begin{figure}[th]
\centerline{
\psfig{file=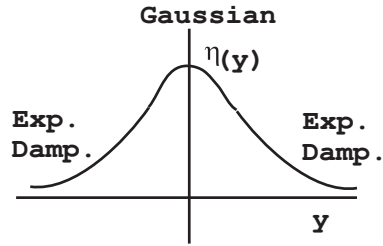,width=5cm}
           }
\vspace*{8pt}
   \caption{
Fig.2\ Behavior of the fermion along the extra axis.
           }
\end{figure}

\section{Five Dimensional QED 
and Bulk Quantum Effect}
Let us examine the 5D QED, 
$\Lcal^{QED}=-e{\bar \psihat}
\ga^\m\ehat_\m^{~m}\psihat A_m$, 
with the Yukawa interaction in RS geometry.
\begin{eqnarray}
\sqrt{-\gh}(\Lcal^{Dirac}+\Lcal^{QED}+\Lcal^Y)\nn
=\sqrt{-\gh}\left[ i{\bar \psihat}
\left\{
\ga^\m\ehat_\m^{~m}(\pl_m+ieA_m)
+\frac{1}{8}(\omhat^\si)_\mn\ga_\si [\ga^\m,\ga^\n]
\right\}\psihat+ig_Y{\bar \psihat}\psihat\Phi
                \right]
\pr
\label{qed1}
\end{eqnarray}
The kinetic (propagator) part for
the electromagnetic, gravitational and Higgs fields is 
provided by 
\begin{eqnarray}
\sqrt{-\gh}(\Lcal^{EM}+\Lcal^{grav}+\Lcal^S)\com\q
\Lcal^{EM}=-\fourth \gh^{mn}\gh^{kl}F_{mk}F_{nl}
\com
\label{qed2}
\end{eqnarray}
where $\Lcal^{grav}$ and $\Lcal^S$ are given in (\ref{yuka2}). 
We assume, as in Sect.4 and 5, $\gh^{mn}$ and $\Phi$ are
the brane background fields obtained as the stable
solution of the system $\Lcal^{grav}+\Lcal^S$.

We have introduced the Yukawa coupling in order to localize
the fermion on the wall. This model, however, is still 
unsatisfactory in that 
the vector (gauge) field is {\it not localized}\cite{BG99}. One resolution
is to take 6D model\cite{Oda00}. Here we are content only with
the fermion part and do not pursue this 
problem.

Let us examine the {\it bulk quantum} effect. It induces the 5D 
effective action $S_{eff}$ which reduces to a 4D action
in the {\it thin wall} limit. From the diagram of Fig.3, we expect
\begin{eqnarray}
\frac{\del S^{(1)}_{eff}}{\del A^\mu(X)}\equiv
<J_\m>\sim e^2g_Y\ep_{\mn\ls\tau}\Phi F^{\nu\la}F^{\si\tau}
\pr
\label{qed3}
\end{eqnarray}
\begin{figure}[th]
\centerline{
\psfig{file=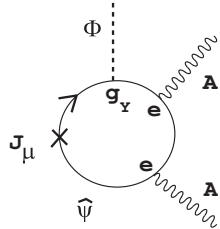,height=3cm}
           }
\vspace*{8pt}
   \caption{
Fig.3\ A bulk quantum-loop diagram. The diagram induces the effective action
$S^{(1)}_{eff}$.
          }
\end{figure}
Then the effective action is integrated as
\begin{eqnarray}
S^{(1)}_{eff}\sim
e^2g_Y\int d^5X\ep_{\mn\ls\tau}\Phi A^\m F^{\nu\la}F^{\si\tau}
\pr
\label{qed4}
\end{eqnarray}
In the {\it thin wall limit} we may approximate as
$\Phi=\Phi(y)\sim \vz\ep(y)$ where $\ep(y)$ is the step function.
(See the description below (\ref{yuka4}).)
Under the U(1) gauge transformation $\del A^\m=\pl^\m\La$, 
$S^{(1)}_{eff}$ changes as
\begin{eqnarray}
\del_\La S^{(1)}_{eff}\sim
e^2g_Y\vz\int d^5X\ep_{\mn\ls\tau}\ep(y) \pl^\m\La
F^{\nu\la}F^{\si\tau}\nn
=e^2g_Y\vz\int d^5X\{
\pl^\m(\ep_{\mn\ls\tau}\ep(y) \La F^{\nu\la}F^{\si\tau})
-\ep_{5\nu\ls\tau}\del(y) \La F^{\nu\la}F^{\si\tau}
                    \}\nn
=-e^2g_Y\vz\int d^4x
\La(x) F^{\ab}{\tilde F}_{\ab}
\com
\label{qed5}
\end{eqnarray}
where ${\tilde F}_{\ab}\equiv \ep_{\ab\ga\del}F^{\ga\del}$. 
In the above we assume that the boundary term vanishes. 
Callan and Harvey interpreted this result as the "anomaly flow"
between the boundary (our 4D world) and the bulk\cite{CH85}. 
Through the analysis of 
the {\it induced action} in the bulk,
we can see the {\it dual} aspect of the 4D QED. 

Another interesting bulk quantum effect is given by Fig.4.
\begin{figure}[th]
\centerline{
\psfig{file=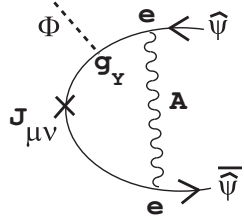,height=3cm}
           }
\vspace*{8pt}
   \caption{
Fig.4\ A bulk quantum-loop diagram. The diagram induces the effective action
$S^{(2)}_{eff}$.
            }
\end{figure}
The induced effective action $S^{(2)}_{eff}$ is expected
to satisfy
\begin{eqnarray}
\frac{\del S^{(2)}_{eff}}{\del F^\mn}\equiv
<J_\mn>\sim e^2g_Y\ep_{\mn\ls\tau}\pl^\la\Phi 
{\bar \psihat}\Si^{\si\tau}\psihat
\pr
\label{qed6}
\end{eqnarray}
Then $S^{(2)}_{eff}$ is obtained as, in the thin wall limit, 
\begin{eqnarray}
S^{(2)}_{eff}
\sim e^2g_Y\ep_{\mn\ls\tau}\int d^5X\pl^\la\Phi\,F^\mn
{\bar \psihat}\Si^{\si\tau}\psihat\nn
= e^2g_Y\vz\ep_{\ab\gd}\int d^4x F^\ab
{\bar \psi}\si^{\gd}\psi
= -ie^2g_Y\vz\int d^4x F^\ab
{\bar \psi}\gago\si_{\ab}\psi
\pr
\label{qed7}
\end{eqnarray}
This term is the EDM term of
(\ref{ferKK5}). The coupling
depends on the vacuum expectation value of $\Phi$, 
$\vz=<\Phi>$ which are, at present, not known.
In order to estimate the magnitude of the
coupling, it is necessary to apply this model
to the quark-lepton (electro-weak) theory and fix
the value. We expect the magnitude could be
sufficiently large so that the result can be tested
by present or near-future experiments.

The appearance of the EDM term corresponds
to the CP-extremely-violated case (ii) $|\mhat|\gg \m$ of Sec.3. 
We compare the parameter relation with the thin wall relation
(\ref{yuka4b}) $k\gg 1/r_c$. 
Because the parameter $\m$ in KK case corresponds to $1/r_c$ in RS case,
we notice the 5D fermion mass in KK case ($|\mhat|$) corresponds to
the inverse of the thickness in RS case ($k$). 

The more fascinating view on the correspondence
is that 
{\it the RS approach and the KK one are "dual" each other}. 
We compare the above thin wall limit
with the small radius limit (i) $|\mhat|\ll \m$ of Sec.3. 
The thin wall limit, which is regarded as the dimensional reduction,
can be consistently taken in RS model and EDM term naturally appears
there. CP is {\it not preserved}. 
The theoretical treatment is justified as far as the relations
(\ref{yuka4b}) and (\ref{yuka4c}) are satisfied:\ 
$1/\sqrt[3]{G_5}\gg k\gg 1/r_c$. 
While, in the KK case, the dimensional reduction takes place
in the case (i) of Sec.3, 
which can be controlled theoretically as far as
the relation (\ref{ferKK8b})
$1/\sqrt[3]{G_5}=\sqrt[3]{100}\m\geq\m\gg|\mhat|$
is satisfied. MDM term
naturally appears and CP is {\it preserved} here.

In the effective action evaluation, the bulk Higgs field
$\Phi(y)$ plays an important role. It serves as a 
bridge between the bulk world and the 4D world.
\section{Discussion and Conclusion}
Since the appearance of EDM in the Randall-Sundrum model was pointed out 
in Ref.\cite{KEK01,SI02PR}, 
several years have passed. The difficulty of the analysis is due to
the lack of the proper treatment of the bulk quantum effects. (We do not have
the renormalizable higher dimensional theory.) Practically, however, an effective
approach called "spurion analysis" has been developed. In Ref.\cite{APS0406,APS0408}, 
the neutron EDM is estimated, in the RSII model, using the approach. They
find the estimated value is 20 times larger than the current experimental
bound and it reveals a "CP problem".  

The present approach to this bulk quantum theory is
different from the above one. We take
the {\it induced effective action} method. It has been
used in relation to  
the chiral and Weyl anomalies.
Famous successful ones are 2D WZNW model derived
from the 2D QCD \cite{PW83PLB}
and 2D induced quantum gravity\cite{Pol87MPL}.
We have examined mainly the thin wall limit.
In order to examine the configuration off the limit,
we can take some numerical approach.

We have comparatively examined the KK model and the RS model.
Both have attractive features as the higher dimensional
unification models. 
The periodic functions appear in the former case, 
while the Bessel functions characteristically appear
in the latter case. The dual property is controlled by
two scale parameters, $\mhat$ and $\mu$ in the KK case
whereas $k$ and $r_c$ in the RS case.
In particular we stress that, as was pointed out by 
Thirring for the KK case, the CP-violation term naturally
appears also in the RS model.

Finally we list the correspondence in Table 1.
\nl
\nl

\begin{tabular}{|c|c|c|}
\hline
             & Kaluza-Klein           & Randall-Sundrum               \\
\hline
electric charge  & $e=f\m$,\ $f$:free para. & $e$:free para. \\
\hline
U(1) sym.  & $y\ra y+\La(x),$ transl. in \{y\}& 
                       $A_m\ra A_m+\pl_m\La,$ internal sym. \\
           & $A_a(x)\ra A_a(x)+\frac{1}{f}\pl_a\La$ & $(X^m)=(x^a,y)$:\ fixed\\
\hline
asym. geometry  & $S^1\times {\cal M}_4$ & AdS$_5$\ $(\om,\mtil=g_Y\vz)$ 
                                                    \\
\hline
           & a massive KK mode for  & 5D bulk Higgs vacuum      \\
           & charged fermion $\psi$,  &   $<\Phi>=\pm\vz, y\ra\pm\infty$ \\
vacuum     & 0-th KK modes for  &  kink sol., $Z_2$-symmetry       \\
           & $g_{ab}, A_a, \si$ and &           \\
           & neutral fermion          &             \\
\hline
4D fermion mass & $\sqrt{\mhat^2+\m^2}$ & $g_Y\vz\times$overlap-int.\\
\hline
physical scale & $\mhat$\ :5D fermion mass  & $k$\ :(thickness)$^{-1}$ \\
\hline
global size  & $\m^{-1}$\ :radius of extra $S^1$    &  $r_c$\ :IR cutoff   \\
             & $y\ra y+2\pi\m^{-1}$ , periodic & $-r_c\leq y\leq r_c$      \\
\hline
dimensional       & $\m\gg\mhat$        &  $k\gg 1/r_c$     \\
reduction cond.   & small radius limit  & thin-wall limit   \\
\hline
5D classical      & $1/\sqrt[3]{G_5}\sim \sqrt[3]{100}\m\gg\mhat$ 
                                        & $1/\sqrt[3]{G_5}\gg k$  \\
condition         &                     &   \\
\hline
mode functions   & $\e^{ik\m y},k\in {\bf Z}$ 
                               & $J_\n(m_kz),N_\n(m_kz),\n=|\half-\mtil/\om|$\\
in extra space   & periodic func. 
                             & $z\om=\e^{\om y}$, $k\in {\bf Z}$,Bessel func.\\
\hline
CP property      & MDM in small radius   & EDM in thin wall   \\
                 & limit, CP-preserved   & limit, CP-violated  \\
\hline
\multicolumn{3}{c}{\q}                                                 \\
\multicolumn{3}{c}{Table 1\ \ Comparison of 
KK model and RS model.  }\\
\end{tabular}


\vs 1

\end{document}